\documentclass[manuscript)]{aastex}

\usepackage{lscape,epsfig,natbib,color}
\usepackage{graphicx}
\usepackage{epstopdf}

\slugcomment{Submitted to ApJ}

\shorttitle{High resolution observations of sympathetic eruptions by NVST}

\shortauthors{Li et al.}

\begin{document}
\title{High-resolution Observations of Sympathetic Filament Eruptions by NVST}

\author{Shangwei Li\altaffilmark{1,2}, Yingna Su\altaffilmark{1}, Tuanhui Zhou\altaffilmark{1}, Adriaan van Ballegooijen\altaffilmark{3}, Xudong Sun\altaffilmark{4}, and Haisheng Ji\altaffilmark{1}}

\affil{$^1$ Key Laboratory for Dark Matter and Space Science, Purple Mountain Observatory, CAS, Nanjing 210008, China;\email{ynsu@pmo.ac.cn}}
\affil{$^2$ University of CAS, Beijing 100049, China}
\affil{$^3$ 5001 Riverwood Avenue, Sarasota, FL 34231, USA}
\affil{$^4$ W. W. Hansen Experimental Physics Laboratory, Stanford University, Stanford, CA 94305, USA;}

\begin{abstract}

We investigate two sympathetic filament eruptions observed by the New Vacuum Solar Telescope (NVST) on 2015 October 15. The full picture of the eruptions is obtained from the corresponding SDO/AIA observations. The two filaments start from active region NOAA 12434 in the north and end in one large quiescent filament channel in the south. The left filament erupts firstly, followed by the right filament eruption about 10 minutes later. Clear twist structure and rotating motion are observed in both filaments during the eruption. Both eruptions are failed, since the filaments firstly rise up, then flow towards the south and merge into the southern large quiescent filament. We also observe repeating activations of mini filaments below the right filament after its eruption. Using magnetic field models constructed based on SDO/HMI magnetograms by flux rope insertion method, we find that the left filament eruption is likely to be triggered by kink instability, while weakening of overlying magnetic fields due to magnetic reconnection at an X-point between the two filament systems might play an important role in the onset of the right filament eruption.  

\end{abstract}

Online-only material: animations, color figures

\section{Introduction} \label{s-intro}

Solar prominences are long thin threads of relatively dense cool gas suspended in the million degree corona. They are observed as ``filaments" on the solar disk, where they are seen in absorption in strong spectral lines (such as H$\alpha$) and in the Extreme Ultraviolet (EUV) continuum. When described as ``prominences" they are seen above the solar limb, where they appear as bright features against the dark background \citep{1985SoPh..100..415H,2010SSRv..151..243L, 2010SSRv..151..333M}. We use the terms ``filament" and ``prominence" interchangeably in general. A filament is formed in a filament channel, which is defined as a region in the chromosphere surrounding a magnetic polarity inversion line (PIL) \citep{1955ApJ...121..349B}.

Solar flares, filament eruptions, and CMEs are thought to be powered by the release of magnetic free energy, which stores in non-potential coronal magnetic fields. There are mainly two groups of models for the magnetic field configuration prior to the eruption. One group assumes that there is a twisted flux rope above the PIL before eruption \citep{Forbes1991,Gibson1998,Krall2000,Wu1997,Roussev2003}. Another model claims that pre-flare configuration consists of sheared arcades, and a twisted flux rope is formed during eruption due to magnetic reconnection \citep{mikic1994,antiochos1999,amari2003,manchester2003}.  

Two groups of mechanisms have been proposed to trigger filament eruptions. One group requires magnetic reconnection, which might be due to magnetic flux emergence \citep{chen2000} or flux cancellation \citep{van1989}, as well as the breakout model \citep{antiochos1999}. Another group is the ideal MHD instability, which includes the helical kink instability \citep{williams2005,rust2005,2007ApJ...668.1232F,2009ApJ...703.1766S} and torus instability \citep{kliem2006,kliem2014} and/or loss of equilibrium \citep{forbes1990,lin2000}. A force free cylindrically symmetric twisted flux tube of infinite length is shown to be always unstable to the helical kink instability \citep{1968SoPh....3..298A}. Considering cylindrically symmetric force free magnetic flux tubes line-tied at both ends, \citet{1981SoPh...73..289H} show that for a uniformly twisted flux tube the kink instability sets in if  the twist angle of the field lines in going from one end to the other exceeds 2.49$\pi$, or about 1.25 full rotations. The twist angle highly depends on the configuration of the flux rope and the ambient coronal field. Models of flux ropes in an external field give values closer to 3.5$\pi$ for the critical twist \citep{2003ApJ...589L.105F, 2004ApJ...609.1123F, 2004A&A...413L..27T}. This value even increases with increasing aspect ratio of the loops involved \citep{2001A&A...367..321B, 2004A&A...413L..27T}. On the other hand, the torus instability is an expansion instability associated with a toroidal current ring held in equilibrium in an external potential magnetic field \citep{kliem2006, 2010ApJ...718.1388D}. The threshold of torus instability is given in terms of the decay index of the external poloidal field, at the position of the  current channel, $n = -\partial{lnB}/\partial{lnz}>n_{cr}$. The critical index ($n_{cr}$) for torus instability ranges from 1.0 to 2.0 in theoretical calculations and numerical simulations \citep{kliem2006, 2007ApJ...668.1232F, 2010ApJ...708..314A, 2010ApJ...718.1388D}.  

To determine what triggers filament eruptions and how the energy is released, it is important to understand the three-dimensional (3D) structure of the coronal magnetic field prior to the eruption. Direct measurement of the coronal magnetic fields is challenging, although recently some progress has been made \citep[e.g., ][]{1998ApJ...500.1009J, 2003Natur.425..692S, 2004ApJ...613L.177L}. Therefore, extensive studies have been pursued to extrapolate the coronal magnetic fields using the photospheric vector magnetograms as the bottom boundary condition \citep[e.g., ][ and references therein]{2008JGRA..113.3S02W,2014A&ARv..22...78W}. In some cases, the reconstructions of the coronal field match observations \citep[e.g., ][]{schrijver2008a,canou2010,yelles2012,valori2012}. But in other cases, the extrapolations have difficulty in producing reliable results \citep[e.g., ][]{metcalf2008,derosa2009}. Alternative technology has also been developed, in particular, the ``flux rope insertion method" \citep{van2004}. In this method, nonlinear force-free field models are constructed by inserting magnetic flux ropes into a potential-field model of the active region and then applying magneto-frictional relaxation \citep{yang1986,van2000}. This method has been used by \citet{bobra2008} and \citet{su2009a,su2009b,su2011} to study active region filaments, and polar crown filaments \citep{2012ApJ...757..168S, 2015ApJ...807..144S}, as well as sigmoids \citep{2009ApJ...703.1766S,2015ApJ...810...96S}.

In this work, we investigate the high resolution observations of two small failed, sympathetic filament eruptions. Detailed analysis of the observations is presented in Section 2. In order to understand the magnetic field structure supporting the filaments as well as the trigger mechanism of the failed eruptions, we perform magnetic field modeling of the region, which is shown in Section 3. Summary and discussions are provided in Section 4.

\section{Observations} \label{s-obs}

\subsection{Data sets and instruments} \label{s-instru}

The primary dataset for this study is taken by the $\emph{New Vacuum Solar Telescope}$ \citep[NVST,][]{2014RAA....14..705L}. The cadence of H$\alpha$ observations is about 49 s and the pixel size is 0.$^{\prime\prime}$167, with a field view (FOV) of 170$^{\prime\prime}$ $\times$ 170$^{\prime\prime}$. Corresponding multi-wavelength images and vector magnetograms taken by  the Atmospheric Imaging Assembly \citep[AIA, ][]{2012SoPh..275...17L} and the Helioseismic and Magnetic Imager \citep[HMI, ][]{2012SoPh..275..229S} aboard  the $\emph{Solar Dynamics Observatory}$ (SDO) are also used for the study. The pixel size of AIA is 0.$^{\prime\prime}$6, and the cadence is 12 s. The HMI magnetograms have a pixel size of 0.$^{\prime\prime}$5 and a cadence of 45 s. Soft X-ray light curve from GOES is also included.

\begin{figure}[ht!]
\plotone{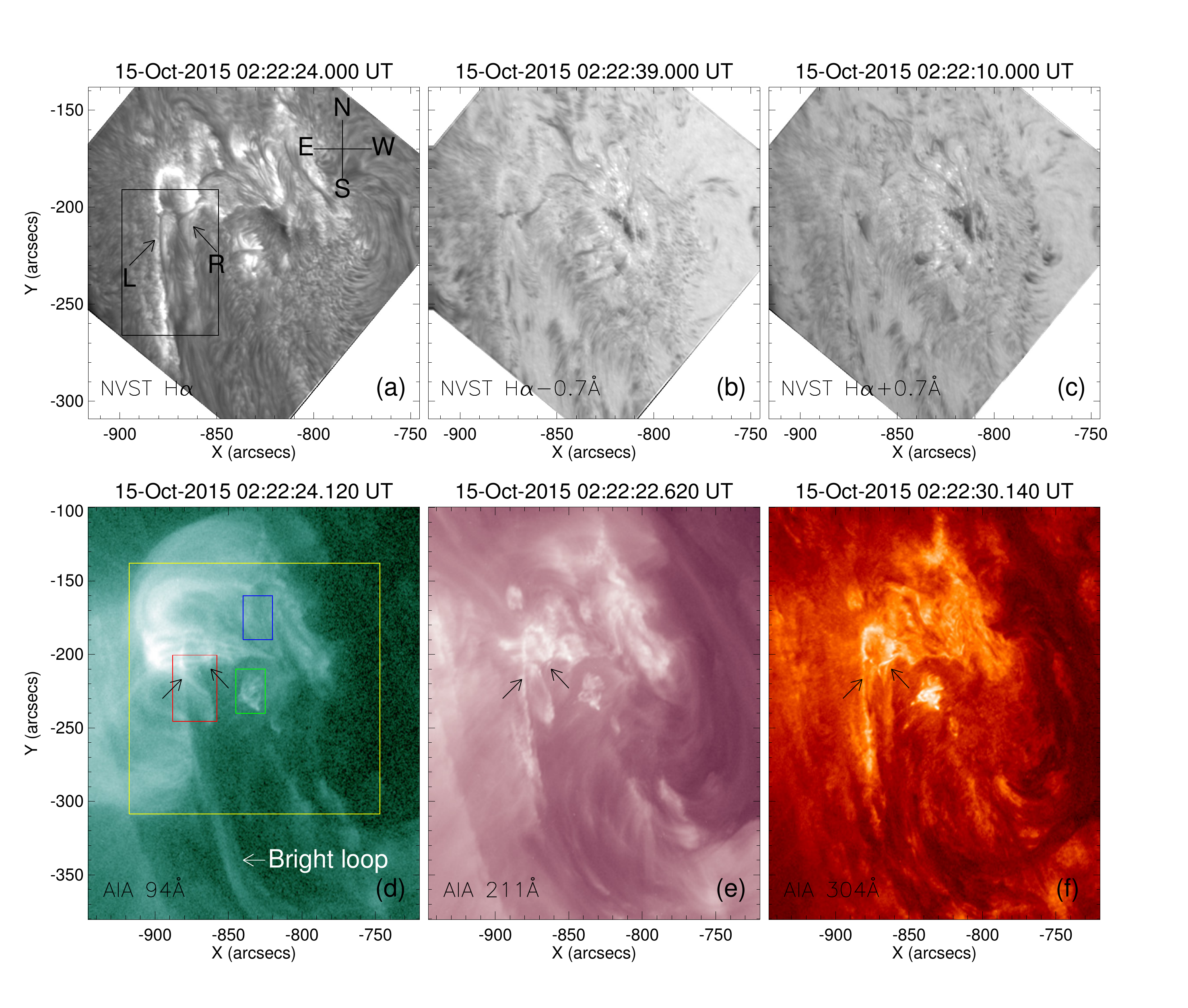}
\caption{NVST and SDO/AIA images of AR 14234 before eruption. (a--c) NVST images at line center, blue and red wings of H$\alpha$ at 02:22 UT. (d)--(f) AIA EUV images in 94~\AA, 211~\AA, 304~\AA~at 02:22 UT. The yellow box represents the FOV of NVST. The left and right filaments are marked with black arrows. \label{fig1}}
\end{figure}

Active region (AR) NOAA 14234  is observed in line center and off-bands ($\pm$0.7~\AA) of H$\alpha$ 6562.8 {\AA} by NVST from 01:59:52 UT to 04:39:23 UT on 2015 October 15. NVST and AIA images of this region before the eruption are shown in Figure~\ref{fig1}. There are two small filaments (left and right) starting from AR 12434 in the north and ending in one large quiescent filament channel in the south. The fine structure of the northern two filaments (marked by black arrows) is best seen in the high resolution H$\alpha$ images (top row). The left filament appears to be straight and narrow, while the structure of the right filament is not clear before eruption. The southern large quiescent filament is shown in the corresponding AIA images with larger field of view (FOV, bottom row). A long bright loop structure is visible in 94~\AA~in the southern quiescent filament channel (Figure 1d), and it appears to be dark in other AIA channels (e.g., Figures 1e--1f). The appearance of this bright loop is caused by a GOES C1.5 class flare occurred at 01:22 UT.  

\subsection{Structure and dynamics of filament eruptions} \label{s-dynamics}

Flares and filament activations in different locations of this region appear to occur in association with each other. To understand the temporal relation between each other, light curves of different areas (marked in Figure 1d) from 02:00 UT to 06:00 UT on 2015 October 15 are plotted in Figure 2.  The peak of the left filament eruption is around 03:45 UT (brown dashed line), while the second eruption peaks at 04:01 UT or so (black dashed line). The peak of the left filament eruption is stronger than the right one (red curve). For the left filament eruption, corresponding peaks are identified in the light curves of northern part (blue) and circular ribbon region (green), but no peak is identified in association with the right filament eruption. These results are consistent with the fact that the left peak is much stronger than the right one in the corresponding GOES light curve (Figure 2a) which refers to the intensity of the full solar disk. We focus on the left and right filament eruptions in this study.

In order to investigate the fine structure and dynamics of the left filament eruption, we present series of NVST and AIA images with a small FOV (black box in Figure 1a) in Figure 3, and corresponding animation Video 1 is presented online. The left and right filaments are marked with letters ``L" and ``R". The top row of Figure 3 shows images in H$\alpha$ at different time, and corresponding AIA images in 304~\AA, 211~\AA, and 94~\AA~ are presented in the second to fourth rows, respectively. Figure 3a1 shows that the pre-eruption left filament appears to be narrow and dark in H$\alpha$, and the fine structure is difficult to identify at 02:55 UT. No clear filament structure can be identified in the corresponding AIA images. Around 03:01 UT, small brightenings appear right next to the left filament as shown in Figure 3b1, and the size of the brightenings is larger in AIA (Figures 3b2--3b4). The bright area grows much larger and the left filament becomes much wider, and signature of twist begins to appear at 03:25 UT. In the corresponding AIA images, brightenings are located at the two ends of the dark filament, and a bright arc-like structure becomes visible in 94~\AA. The apparent right-handed twist structure becomes much clearer in H$\alpha$ as the dark filament becomes wider, and brighteings spread in a much larger area at 03:31 UT. Figure 3d1 shows that the filament is composed of fine threads which are not parallel to, but at a certain angle with respect to the filament spine. Alternative bright and dark structures are visible in the filament in 304~\AA. Five minutes later, the southern portion of the left filament shows clear rise, and the region becomes mostly saturated in both H$\alpha$ and AIA 304~\AA. During the rise, the southern portion of the left filament displays clockwise rotation (viewing from the South) in H$\alpha$, which can be identified by comparison of Figures 3c1--3e1 and Video 1. The clockwise rotation is likely to be the untwisting of the erupting filament. The arc-like structure becomes much larger and brighter in 94~\AA~at 03:36 UT as shown in Figure 3e4. The brightenings are mostly stronger in 304~\AA~than the other two AIA passbands throughout the eruption.

\begin{figure}[ht!]
\plotone{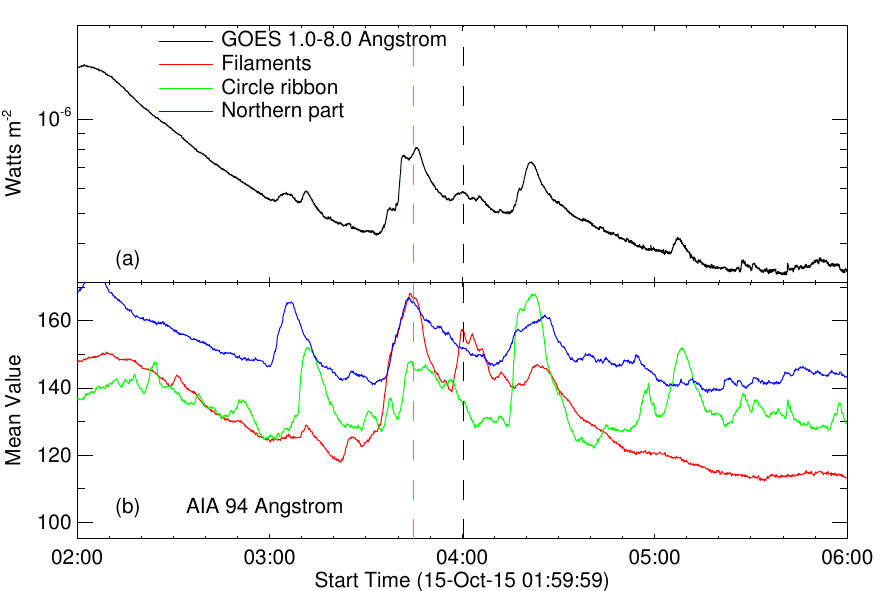}
\caption{(a) GOES X-ray light curve from 02:00 UT to 06:00 UT on 2015 October 15. (b) AIA 94~\AA~light curves of average data numbers of different regions enclosed by colored boxes in Figure 1d. The brown and black vertical dashed lines refer to the peaks of the left and right filament eruption.
\label{fig2}}
\end{figure}

The large scale structure and dynamics of the erupting left filament are shown in a series of AIA images with large FOV in Figure 4, and corresponding animation Video 2 is presented online. The first column shows the locations of the filaments (black arrows) at the onset of the left filament eruption, when we starts to see brightenings. The bright emission propagates in a much larger area as time goes on, and brightenings becomes visible in the southern quiescent filament channel starting around 03:11 UT. These brightenings then continue to propagate towards the south until they reach the southern end of the filament channel (white and yellow arrows in the second and third columns). Around 03:42 UT, a large scale bright loop structure is visible in 94~\AA, and corresponding brightenings can also be identified in 304~\AA~and 171~\AA~(fourth column). This figure and Video 2 show that the southern portion of the left filament firstly rises up, then most of the filament materials flow along the southern filament channel.  Weak flows in the southern part of the erupting left filament before its rise (i.e., before 3:00 UT) are also observed in 304~\AA~(see Videos 1 and 2). The left filament eruption is likely to be failed since no clear filament or magnetic structure eject outwards successfully, and no corresponding coronal mass ejection (CME) is observed. The red, blue and green arrows in Figure 4c4 mark the brightenings at different locations in association with the left filament eruption, using the same colors as for the boxes in Figure~\ref{fig1}d and their light curves in Figure~\ref{fig2}b.

The right filament becomes darker and rises slowly in association with the left filament eruption as shown in the top row of Figure 3. The left filament is a more violent process since bright emissions are observed in many passbands, e.g., H$\alpha$, 304 {\AA}, 211{\AA} and 94 {\AA}. While for the right filament eruption, only significant brightenings are observed in 304 {\AA} as shown in Figure 5, and corresponding animation Video 3 is presented online. From the top to bottom rows, Figure 5 displays a series of images of the right filament eruption in H$\alpha$ by NVST and 304~\AA, 211~\AA, and 94~\AA~by AIA. Significant rise of the right filament begins around 03:45 UT. Brightenings around the right filament appear immediately in H$\alpha$ and 304~\AA~as shown in Figures 5a1--5a2. The right filament grows thicker and displays an arc-like structure at 03:52 UT. Three minutes later, H$\alpha$ image shows appearance of twist signature, and several filament threads are aligned at a certain angle with respect to rather than parallel to the filament spine. Around 03:56 UT,  the apparent right-handed twist structure becomes most clear in H$\alpha$ as shown in Figure 5d1, and it is also visible in 211 {\AA}. Brightenings around the right filament gradually increase as it rises in height, which is mainly observed in 304 {\AA}. By 04:00 UT, most of the erupting filament materials become saturated in 304~\AA, while no significant brightenings are observed in other bands (fifth column). Similar to the left filament, the southern portion of the right filament also displays clear clockwise rotation (untwisting motion) during its rise as revealed by Figure~\ref{fig5}c1--e1 and Video 3, from which we discover the gradual rise of a mini filament under the erupting right filament. The mini filament continues to rise, until it disappears at certain height.  This behavior of the mini filament is best observed in H$\alpha$, and it is also indicated in the corresponding AIA observations (e.g., in 304~\AA~, 211~\AA).

Figure~\ref{fig6} presents a large scale picture of the right filament eruption, and corresponding animation Video 4 is presented online. The large bright loop structure at the onset of the right filament eruption (left column) is due to the left filament eruption. The right filament firstly rises up, then both dark and bright materials flow towards the south (white and yellow arrows, see also Video 4). The flowing filament materials are mostly visible in 304~\AA~and 171~\AA, and only faint emission can be identified in 94~\AA. Therefore, the right filament eruption is also failed, since no clear outward ejection of filament materials and magnetic structure, and no corresponding CME are observed. A comparison of the first and fourth columns of Figure 6 shows that the left and right filaments are flowing along different paths of the southern filament channel towards the south during the eruption.

An overview of the temporal evolution of the two filament eruptions is presented in Figure 7, which shows three time-slice diagrams along different paths marked by S1 (northern portion), S2 (southern portion) and S3 in Figure~\ref{fig5}. Diagrams of S1 and S2 refer to slices cutting from H$\alpha$ images, and S3 is for 304 {\AA} images. Arrows ``L" and ``R" mark the location of the left and right filament. The left filament eruption is better captured in diagram S1, which shows that the left filament begins to rise slowly around 02:00 UT, when slow rise of the southern portion of the right filament also begins as shown in diagram S2. The red dashed line marks the start time (~03:35 UT) of the fast rise or explosive eruption of the left filament, which is associated with clear rise of the right filament (Figure 7a) and brightenings nearby (Figure 7c). Significant brightenings begin to appear below the right filament around 03:45 UT (green dotted line, Figure 7c), when the fast rise of the southern portion of the right filament begins (Figure 7b). As most of the right filament materials rise up, part of them falls back towards the Sun (yellow arrow in Figure 7a). Repeating activations of mini filaments below the right filament during and after its eruption is observed in NVST and most of the AIA EUV channels. Two of the mini filaments' activations are captured by NVST as shown in Figure~\ref{fig7}b. The mini filaments firstly rise up then disappears at certain height. Mini filaments are invisible in Figure 7c since slice 3 is away from the location of the mini filaments.

\section{Magnetic field modeling} \label{s-model}

The northern two small filaments erupt one after another, during which filament threads display right-handed twist structure and clockwise rotating motion, which might be the manifestation of twisted flux ropes. To understand the magnetic structure supporting the erupting filaments as wells as the trigger mechanisms of the eruptions, we construct magnetic field models using the flux rope insertion method developed by van Ballegooijen \citep{van2004}. We briefly introduce the method below, for detailed descriptions please refer to \citet{bobra2008, su2009a, su2011}.

At first, a potential field model is computed from the high-resolution (HIRES) and global magnetic maps observed by SDO/HMI. If the inclination of the magnetic field is large (more horizontal), false polarity change will be created for a region away from the disk center, which is a known problem with line-of-sight (LOS) magnetograms. Since this region is located near the east limb, the lower boundary condition for the HIRES region is derived from the photospheric vector magnetograms obtained at 02:00 UT on October 15. There are several caveats of magnetic fields near the limb: 1) false PIL if the field is not radial; 2) foreshortening due to projection effect (one pixel covers larger area, this is especially problematic near PIL since opposite polarities may coexist in the same pixel); 3) disambiguation may not work well due to the two aforementioned problems. We are more or less confident with this map as we have inspected the AR when it's closer to central meridian, and the polarity appears to be consistent. The longitude-latitude map of the radial component of the magnetic field in the HIRES region is presented in Figure~\ref{fig8}. The HIRES computational domain extends about 29$^{\circ}$ in longitude, 28$^{\circ}$ in latitude, and up to 1.73 $\rm R_\odot$ from the Sun center. The models use variable grid spacing to achieve high spatial resolution in the lower corona (i.e., $0.0005 R_{\sun}$) while covering a large coronal volume in and around the target region. Three paths marked with blue curves are selected according to the locations of the observed filaments. The two short blue paths represent the left and right filaments, which are located in two adjacent bipoles `P1/P2' and `P3/P4'. The longer path is consistent with the long loop structure brightened during eruption, which is located in the quiescent southern region. Next we modify the potential field to create cavities in the region above the selected paths, then insert three thin flux bundles (representing the axial flux $\Phi_{\rm axi}$ of the flux rope) into the cavities. Circular loops are added around the flux bundle to represent the poloidal flux $F_{\rm pol}$ of the flux rope.  The two small filaments appear to be low lying since they are located near or partially rooted in the active region, while large quiescent filaments are higher in altitude in general. Therefore, we insert the longer flux rope in a higher altitude than the two short ones. All of the three inserted flux ropes contain right-handed twist. The resulted magnetic fields are not in force-free equilibrium. We then use the magneto-frictional relaxation to drive the field towards a force-free state \citep{van2000, yang1986}. 

\subsection{Modeling of the filaments} \label{model-left}

After a few attempts, we choose axial fluxes of 3$\times 10^{20}$ and 6$\times 10^{20}$ Mx, poloidal fluxes of 3 $\times 10^{10}$ and $10^{10}$ Mx cm$^{-1}$, for the right filament and the long quiescent filament, respectively. The resulted model matches the observed filaments and coronal loop structures well.  We focus on the left filament since it starts eruption firstly. We build a series of models in order to find the best-fit model representing the observed left filament. At first, we construct four models (Models 1--4) with a fixed poloidal flux of $10^{10}$ Mx cm$^{-1}$, and different axial fluxes of 0, 6, 10, 20 in unit of $10^{20}$ Mx. The model results after 30000-iteration relaxation are shown in Figure~\ref{fig9}. The top row shows distribution of electric currents in a vertical cross section of the flux rope along the yellow slice (starting from the positive polarity region) marked in Figure~\ref{fig8}. The currents of the core fields represented by white regions are increasing in height as the axial flux increases. Selected field lines (color lines) representing the core fields from different models are displayed in the bottom row. The core fields appear to be sheared arcades, and the southern footpoint firstly extends towards the east, then no significant change appears with increasing axial flux. Although Model 4 is more unstable in comparison to Model 1, it does not show clear signature of instability like that shown in \citet{su2011}, for example, appearance of X-point within or below the non-potential core fields and/or continue expansion after 30000-iteration relaxation. In fact, the left filament eruption is not successful. These attempts suggest that it is difficult to reproduce the clear twist structure of the observed filament by increasing the axial flux of the inserted flux rope.

According to the aforementioned attempts on the axial flux, we find that an axial flux of  $3 \times 10^{20}$ Mx for the left flux ropes leads to a good match to the observed length and location of the left filament. Therefore, we then construct four models (Models 5--8) with a fixed axial flux of $3 \times 10^{20}$ Mx, and poloidal fluxes of 0, 5, 10 and 30 in unit of $10^{10}$ Mx cm$^{-1}$, respectively. Similar to Figure ~\ref{fig9}, distribution of electric currents and selected field lines from these models after 30000-iteration relaxation are shown in Figure ~\ref{fig10}.  Model 5 with poloidal flux of 0 Mx/cm turns out to be sheared arcades after relaxation. A flux rope with a current-sheet like structure is developed when the poloidal flux is $5\times10^{10}$ Mx cm$^{-1}$ (Model 6). An X-point like feature appears in the current sheet region below the flux rope for Model 7 with poloidal flux of $10\times10^{10}$ Mx cm$^{-1}$.  The existence of an X-point-like feature below the flux rope supports the suggestion in \citet{Savcheva2012} that this feature indicates that the region has come close to the onset of eruption. Another X-point marked with a yellow arrow in Model 8 shows significant motion towards the west in comparison to the other models, which suggests that the flux rope is expanding significantly. The bottom row of Figure ~\ref{fig10} shows that selected field lines of the flux ropes become more twisted as the poloidal flux increases.

\subsection{Comparison with observations} \label{model-right}

Observations in Section 2.2 have revealed clear twist structure and rotating motion as the filaments rise up. After a detailed comparison, we find that Model 7 shows best fit to the observed filament at the early stage of the eruption. The twist of the core fields in the other models is either too large or too small. A comparison of Model 7 with observations is presented in Figure~\ref{fig11}. The fine structure of thin filament threads in both left and right filaments is relatively well represented by the twisted field lines in Model 7 (left and middle columns). A long weakly twisted flux rope (close to sheared arcades) shows good match to the observed bright long loop-like structure (right column). Discrepancy does exist, for example, for the southern portion of the left filament, and for the height of the right flux rope which appears to be lower than the observed right filament. The discrepancy regarding the height of the right filament is likely to be due to the fact that the observed filament is erupting, while the modeled magnetic field is in a pre-eruption state. The image of the erupting filament is chosen to show the twist structure, which appears to be consistent with the modeled flux rope to some degree. It is difficult to determine the northern footpoint of the long flux rope from observations, so we place it in the region with enough positive magnetic flux. The fact that these two filaments are really small and they are located near the limb should also contribute to the discrepancy. A better model could be constructed if we had more accurate high-resolution magnetic field observations. Due to current limitations, we aim to find a relatively good model that can help us to understand the twist structure revealed in observations as well as the large scale magnetic configuration in this region.

Another interesting phenomenon we observe is the appearance of a bright arc-like structure after the initiation of the left filament eruption, which is followed by eruption of the right filament. To understand these observations, detailed examination of the magnetic configuration in this region is presented in Figure~\ref{fig12}. We select a slice along the yellow line `s' starting from the positive polarity region marked in Figure~\ref{fig12}a. Distribution of the electric currents in a vertical cross section (`s-z' plane) is displayed in Figure~\ref{fig12}b. We find that the altitude of the long flux rope/sheared arcade (FR3) is indeed higher than the other two (FRs 1--2) as we expect. In the current distribution plot, we find an X-point located at a height of 7 Mm above the solar surface. This X-point separates the flux ropes into two systems, i.e., one includes the left flux rope FR1, and the other system is composed of the right flux rope FR2 and the long flux rope FR3. Four selected field lines represent different line systems around this X-point are plotted, and the four white circles refer to the locations where the field lines interact with the `s-z' plane. Figure~\ref{fig12}c shows 3D plots of these four field lines overlaid on one 94~\AA~image by AIA. This dark blue line system (3) appear to be produced by magnetic reconnection at the X-point between the light blue line system (1) above the left filament and the pink line system (2) above the right filament. The bright arc-like structure marked with white arrow in Figure~\ref{fig12}c is located between the two bipoles hosting the two erupting filaments. The observed bright arc and the dark blue field line (3) share similar appearance, suggesting of similar formation mechanism. 

\section{Summary and Discussions} \label{s-sum}

In this paper, we report high resolution observations of two sympathetic filament eruptions by ground-based telescope NVST in AR NOAA 12434 on 2015 October 15. In the northern part, the two small filaments are hosted by adjacent two magnetic bipoles, which are located in AR 12434. Complementary observations by AIA and HMI aboard SDO show a long quiescent filament in the southern bipole. NVST observations show that the slow rise of both filaments with no corresponding brightenings begin no later than 02:00 UT. The left filament evolves from a straight narrow dark filament to a wide structure after the appearance of brigtenings in the surroundings around 03:01 UT. The fine filament threads display apparent right-handed twist structure as it becomes wider. The southern end of the left filament displays a clockwise rotation when looking from the south as it rises up. Afterwards, all of the erupting filament materials appear to flow along the southern quiescent filament towards the south end. The southern filament is brightened up in 304~\AA, and corresponding 94~\AA~observations show a bright long loop like structure. Fast rise of the right filament begins at 03:45 UT, which is about 10 mins after the fast rise of the left one. Similar to the left filament, the right one also displays apparent right-handed twist structure and clockwise rotation during the eruption. Most of the right filament materials are brightened and flow along the southern quiescent filament towards the south, but the long loop structure is very faint in 94~\AA. Light curve and imaging analysis suggest that the right filament eruption is weaker than the left one.

To understand the observed twist structure and dynamics as well as the initiation mechanism of the erupting filaments, we construct  a series of magnetic field models using flux rope insertion method. Three inserted flux ropes with right-handed twist include two small lower ones, and a longer one located in a higher altitude. We find that the observed twisted filament threads cannot be reproduced by models with sheared arcades. The model that best describes the twist structure of the left filament during eruption has axial and poloidal fluxes of  $3 \times 10^{20}$ Mx and $10\times10^{10}$ Mx cm$^{-1}$. The twist angle of the flux rope in this model can be estimated as $\Phi = 2 \pi F_{pol} L / \Phi_{axi} $ \citep{2009ApJ...703.1766S}. The length of the flux rope is around  6.3  $\times 10^{9}$ cm. The derived twist angle is about $4.2 \pi$, which is above the critical value (around $3.5 \pi$) for kink instability in the literature \citep[e.g.,][]{2004A&A...413L..27T}. Note that the twist structure is only observed after the onset of the eruption. One important question is whether this twist is formed before the eruption or produced after the eruption onset. The observations appear to be in favor of the earlier scenario. Before eruption, the filament is located in the dips of a tightly twisted flux rope close to the threshold of instability, which is corresponding to the observed narrow dark filament structure. After a small disturbance, likely due to tether-cutting like reconnection indicated by observed brightenings, the magnetic structure supporting the filament becomes unstable after reaching the threshold of kink instability. Then the structure begins to expands wider and untwist (observed clockwise rotation). The filament material starts flowing along the twisted field lines with untwisting motion, which lead to the appearance of clear twisted filament threads. These results suggest that kink instability is likely to play an important role in the initiation of the left filament eruption.

Another interesting observation is that the fast rise of the right filament begins soon after the left filament eruption, suggesting sympathetic eruptions \citep{2011ApJ...739L..63T}. The right filament is modeled with a flux rope with twist below the threshold of instability. From the modeling, we find an X-point located at a height of 7 Mm above the solar surface between the two small filaments. The rise of the left filament triggers magnetic reconnection at the X-point, indicated by the appearance of the bright arc-like structure between the two filaments. This reconnection weakens the constraints of the overlying fields of the right filament, which results in the eruption of the right filament.

Both filament eruptions are failed, since all of the filament materials appear to flow along the southern quiescent filament after a sudden rise, and no corresponding CME is observed. The southern long quiescent filament and bright loop like structure is modeled with a weakly twisted flux rope or sheared arcades located above the two small filaments. The small filaments firstly rise up, then interact with and are constrained by the overlying large quiescent filaments, which might explain the aforementioned observations. Repeating activations of mini filaments are observed after the right filament eruptions. Mini filaments firstly rise up, then disappear after reaching certain height. Activations of mini filaments below an overlying filament has been reported in the literature before eruptions, \citep[e.g.,][]{2014ApJ...789..133Z, 2016RAA....16....3Y}. These observations imply the complex nature of filaments on the Sun.

\citet{2012ApJ...756...59L, 2014ApJ...792..107K} have reported a ``double-decker" configuration, which have two filaments with same chirality, and one is on top of another. In our case, the northern two small filaments also have the same chirality, though side by side. The southern quiescent filament appears to be on top of these two small filaments, and activations of mini filaments are also observed under the northern right filament after it is eruption. Therefore, this region is composed of multiple filaments with difference scales at various locations and altitudes, likely a multi-decker configuration, suggesting the complexity of magnetic topology. Several activations in this region appear to occur closely in time, and all of them might be sympathetic eruptions, though only two sympathetic filament eruptions are studied in this paper. This work has shown the importance of high resolution observations, which can reveal fine structure and dynamics thus improving our understanding of solar activities.

\acknowledgements
We acknowledge the referee for helpful comments to improve the presentation of the paper. Y.S. acknowledges Drs. Yuhong Fan and Rui Liu for valuable discussions. The H$\alpha$ data used in this paper are obtained with the New Vacuum Solar Telescope in Fuxian Solar Observatory of Yunnan Astronomical Observatory, CAS. SDO is a mission of NASA\rq{}s Living With a Star Program. AIA and HMI data are courtesy of the NASA/\textit{SDO} science teams. This work is supported by the Youth Fund of Jiangsu BK20141043, NSFC 11473071 and 11333009, and the One Hundred Talent Program of CAS.

\clearpage

\begin{figure}[ht!]
\epsscale{.80}
\plotone{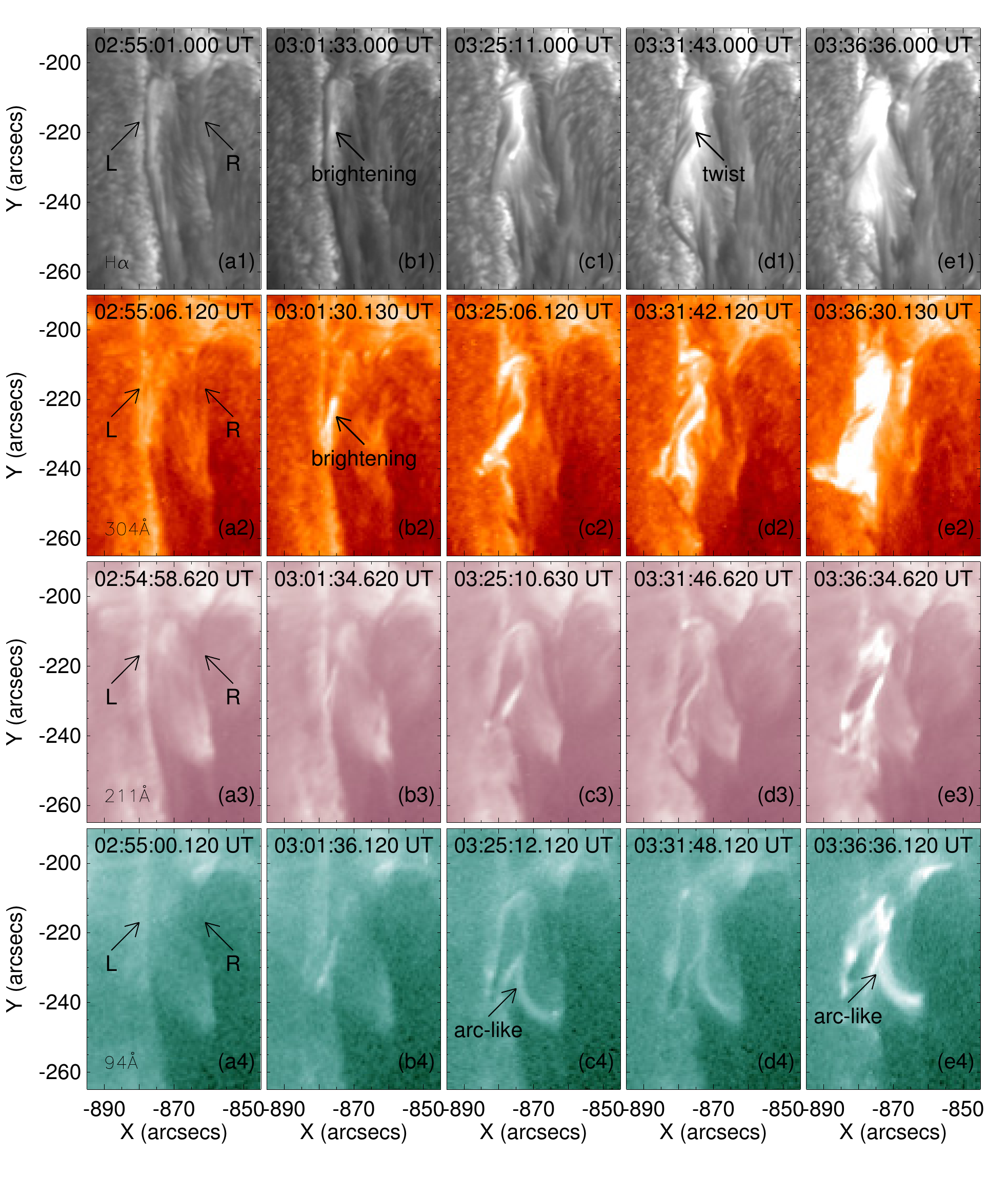}
\caption{Evolution of the left filament eruption in H$\alpha$ by NVST (1st row), 304 {\AA} (2nd row), 211 {\AA} (3rd row) and 94 {\AA} (4th row) by AIA, respectively. The FOV is marked with the black box in Figure~\ref{fig1}a. The arrows L and R mark the location of the left and right filaments. Corresponding animation Video 1 is presented online.
\label{fig3}}
\end{figure}

\begin{figure}[ht!]
\plotone{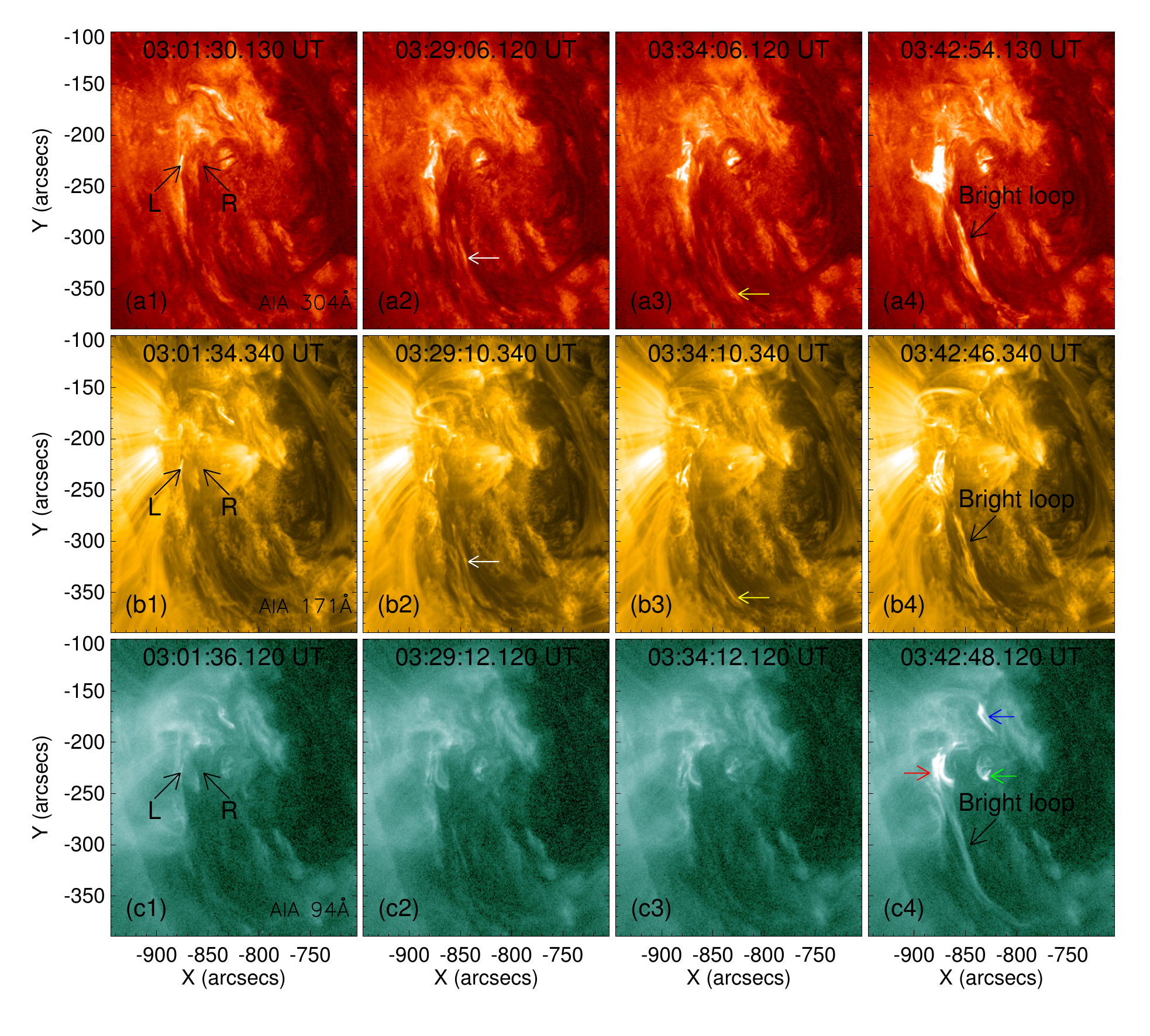}
\caption{Evolution of the left filament eruption in a larger FOV. Images in the first, second and third rows are taken in 304 {\AA}, 171 {\AA} and 94 {\AA} by AIA. The white and yellow arrows in the middle columns refer to materials flowing along the southern filament. The green and blue arrows refer to corresponding activations at various locations marked with boxes having the same color in Figure~\ref{fig1}d.  Corresponding animation Video 2 is presented online.
\label{fig4}}
\end{figure}

\begin{figure}[ht!]
\plotone{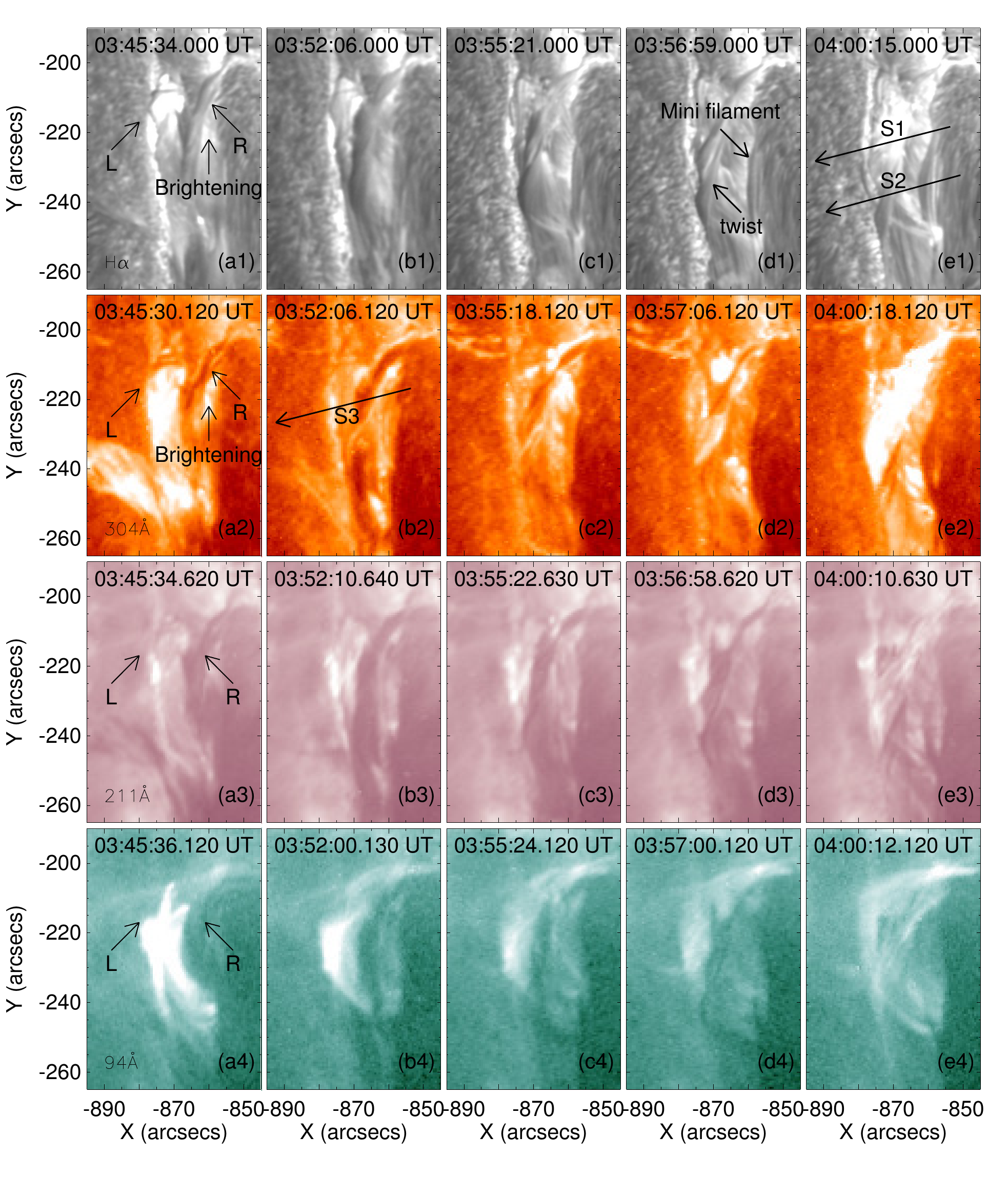}
\caption{Similar to Figure~\ref{fig3}, images of the right filament eruption in  H$\alpha$ by NVST,  304 {\AA}, 211 {\AA}, and 94 {\AA} by AIA are presented in rows 1--4. Corresponding animation Video 3 is presented online.
\label{fig5}}
\end{figure}

\begin{figure}[ht!]
\plotone{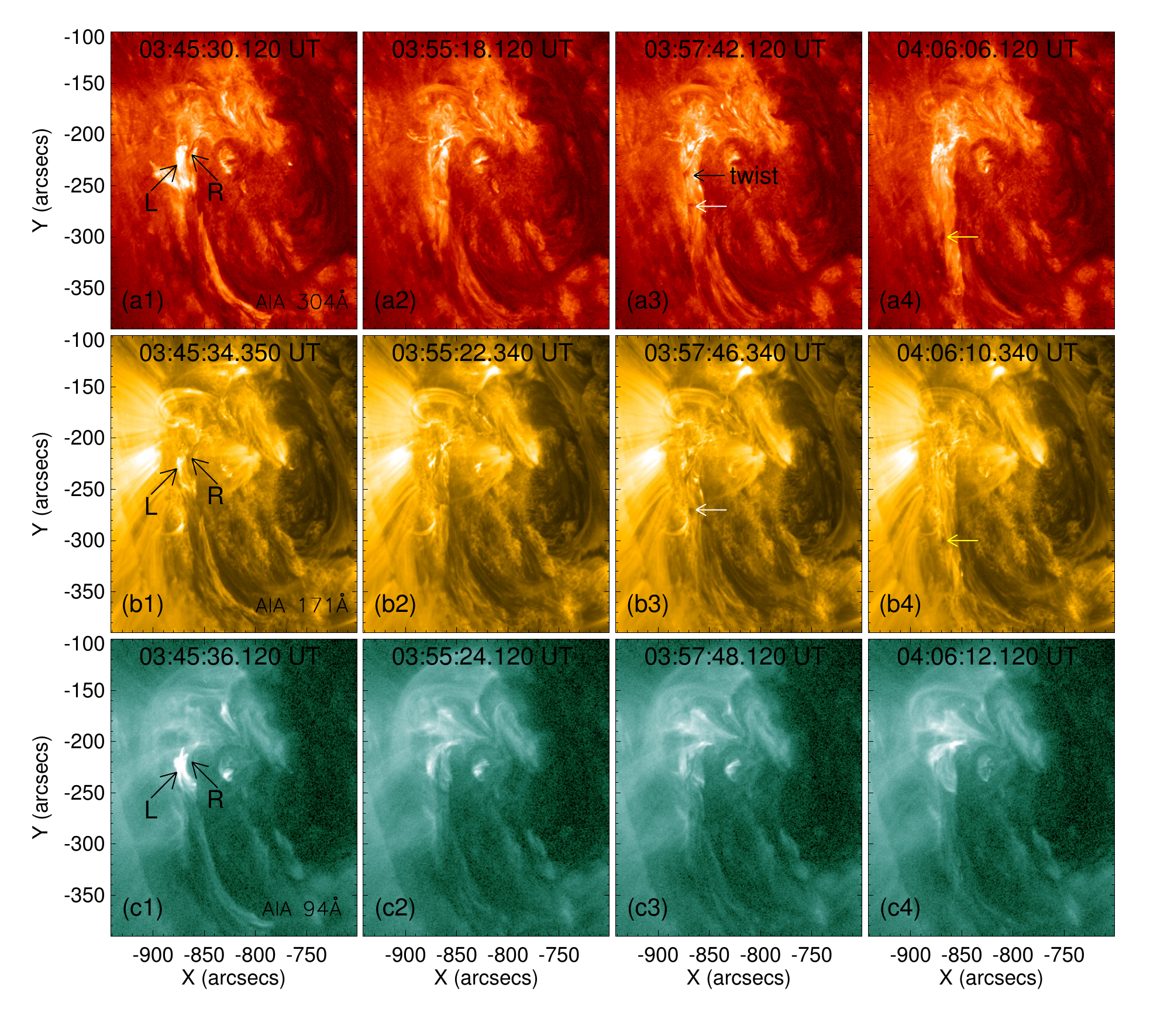}
\caption{Similar to to Figure~\ref{fig4}, but for the right filament eruption. From top to bottom rows, the images are taken in 304 {\AA}, 171 {\AA} and 94 {\AA} by AIA, respectively. Corresponding animation Video 4 is presented online.
\label{fig6}}
\end{figure}

\begin{figure}[ht!]
\plotone{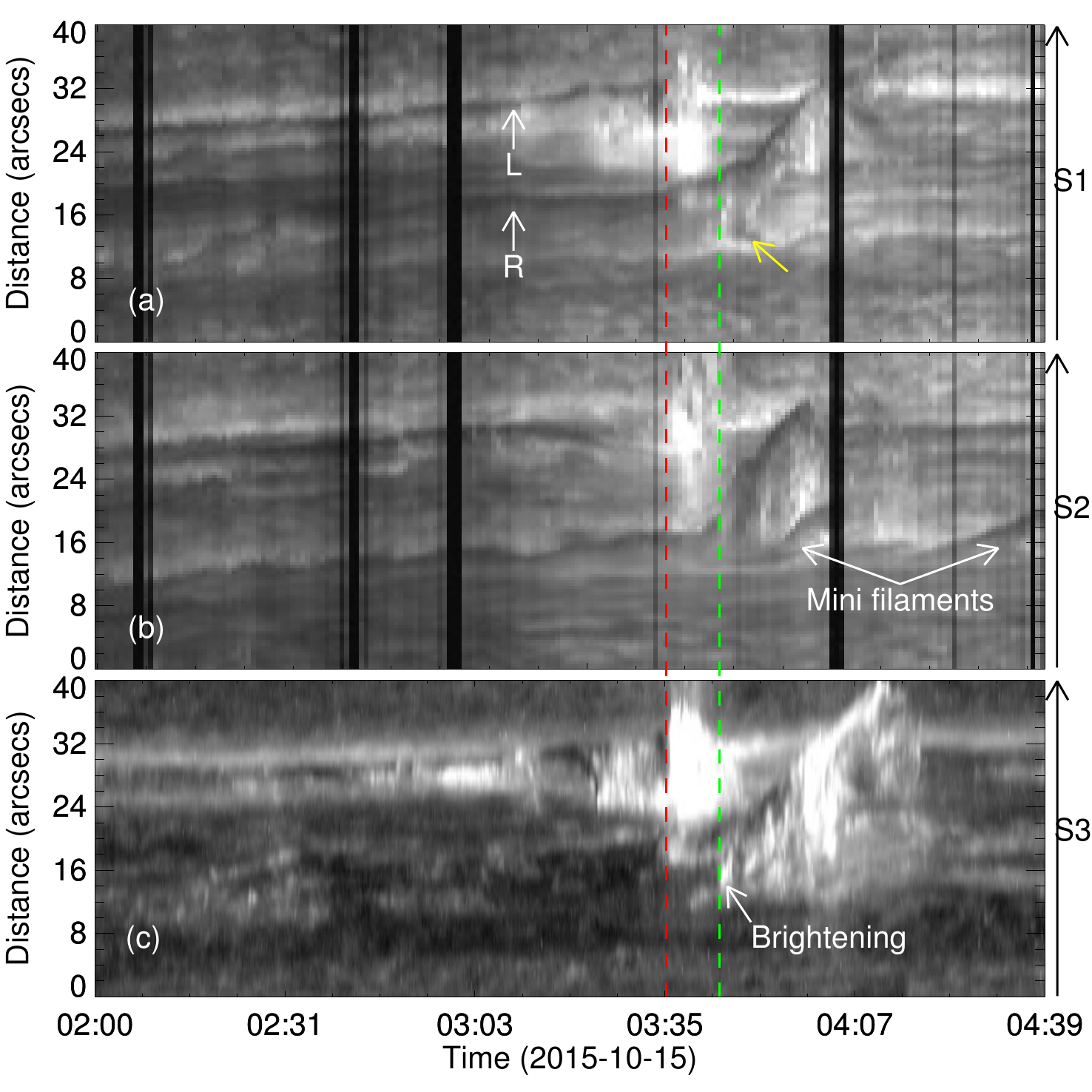}
\caption{(a)--(b) Time-slice diagram of NVST H$\alpha$ images along slices S1 and S2 marked in Figure~\ref{fig5}e1. The arrows L and R represent the location of the left filament and the right filament. The yellow arrow points at the filament materials falling back towards the Sun during the right filament eruption. (c) Time-slice diagram of the SDO/AIA 304 {\AA} images along Slice S3 marked in Figure~\ref{fig5}b2. The red and green dotted lines refer to the timing of significant activations of the left filament eruption, and appearance of significant brightenings below the right filament.
\label{fig7}}
\end{figure}

\begin{figure}[ht!]
\plotone{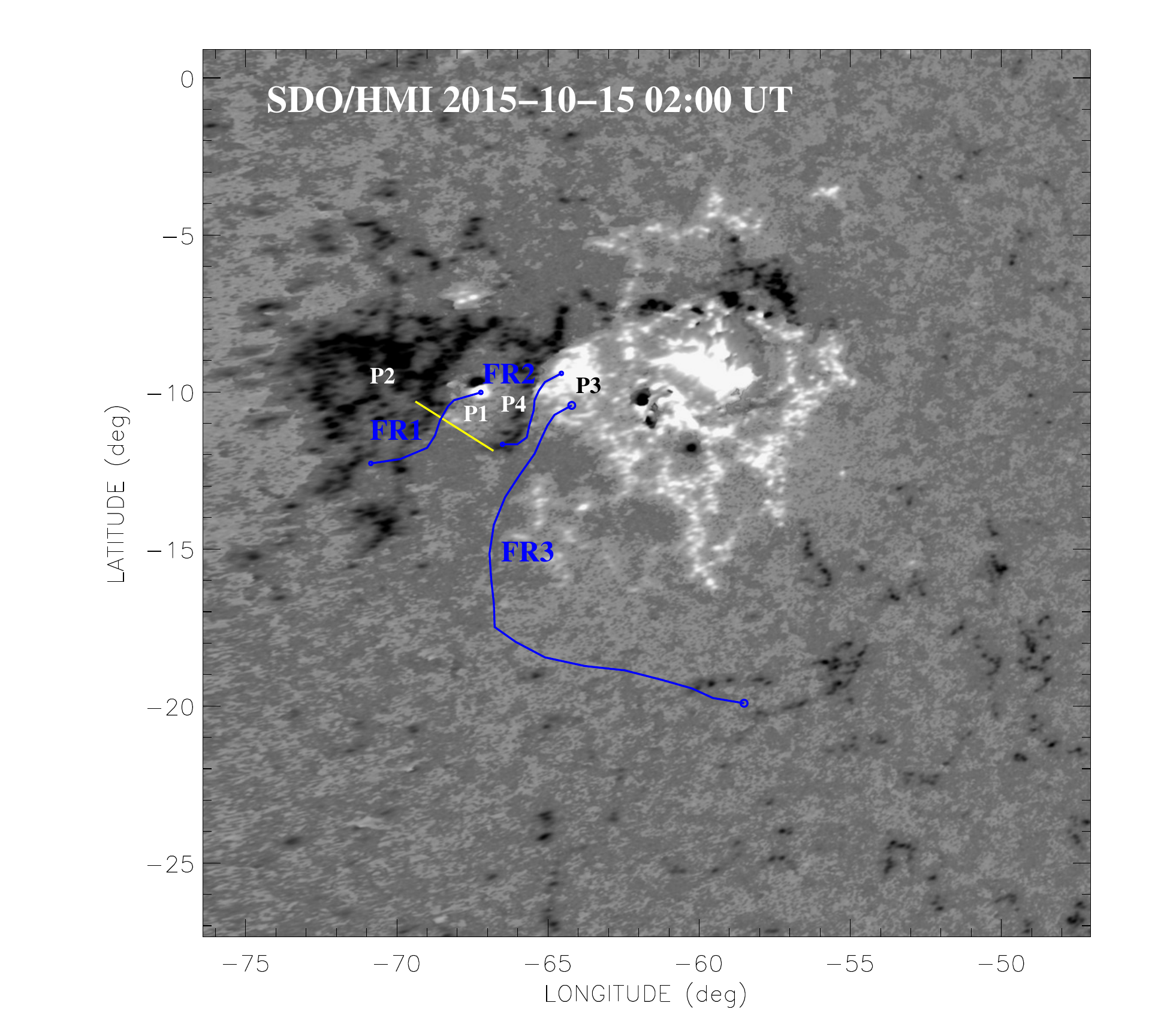}
\caption{The longitude--latitude map of the radial component of the photospheric magnetic field by SDO/HMI in the HIRES region at 02:00 UT on October 15. The blue curves refer to the paths along which we insert flux ropes. 
\label{fig8}}
\end{figure}

\begin{figure}[ht!]
\plotone{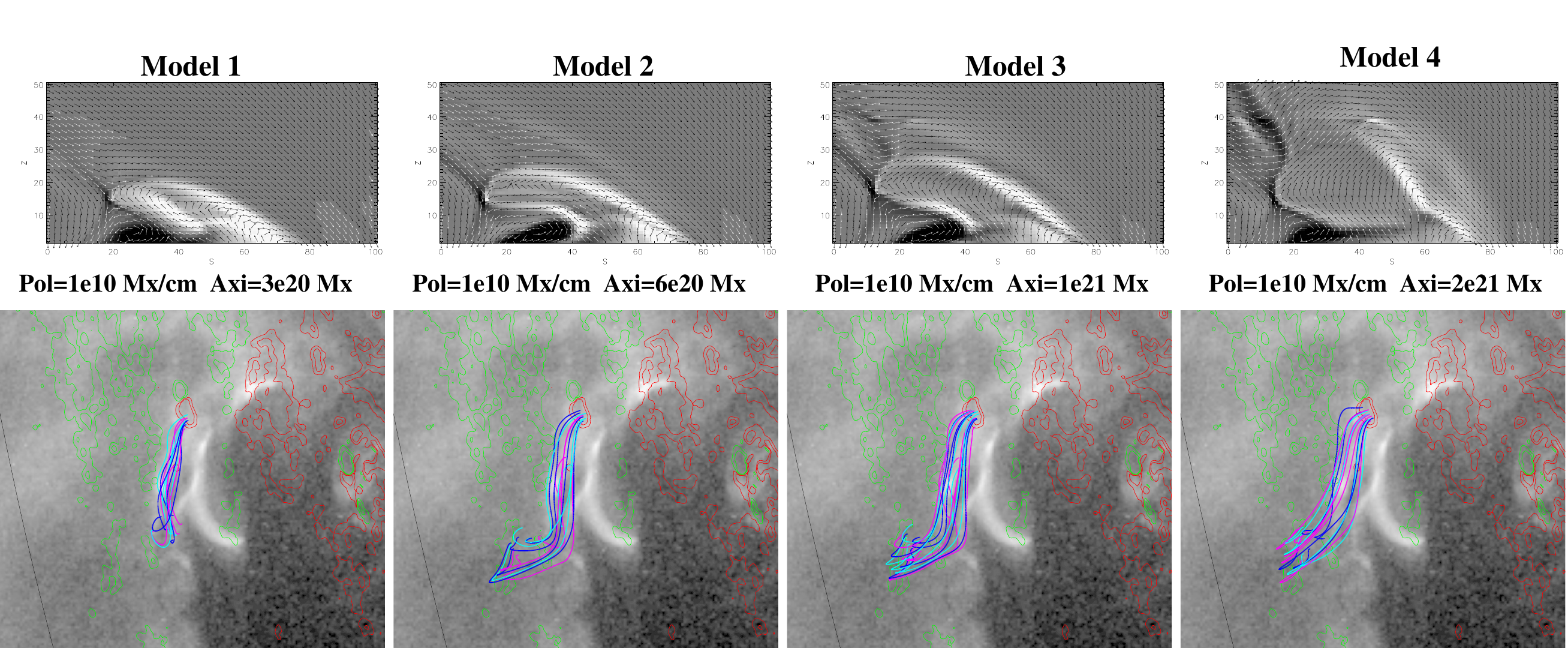}
\caption{Model results for the left filament after 30000-iteration relaxation. The models have fixed poloidal flux and increasing axial flux from left to right. Top row: distribution of electric currents in a vertical cross section along the yellow line marked in Figure ~\ref{fig8}. The
maximum value of the current density is 10000 G $R^{-1}_{\sun}$.  Magnetic vectors are represented by black and white arrows. Bottom row: selected model field lines overlaid on AIA 94 {\AA} images. Red and green contours represent positive and negative magnetic polarities taken by HMI. 
\label{fig9}}
\end{figure}

\begin{figure}[ht!]
\plotone{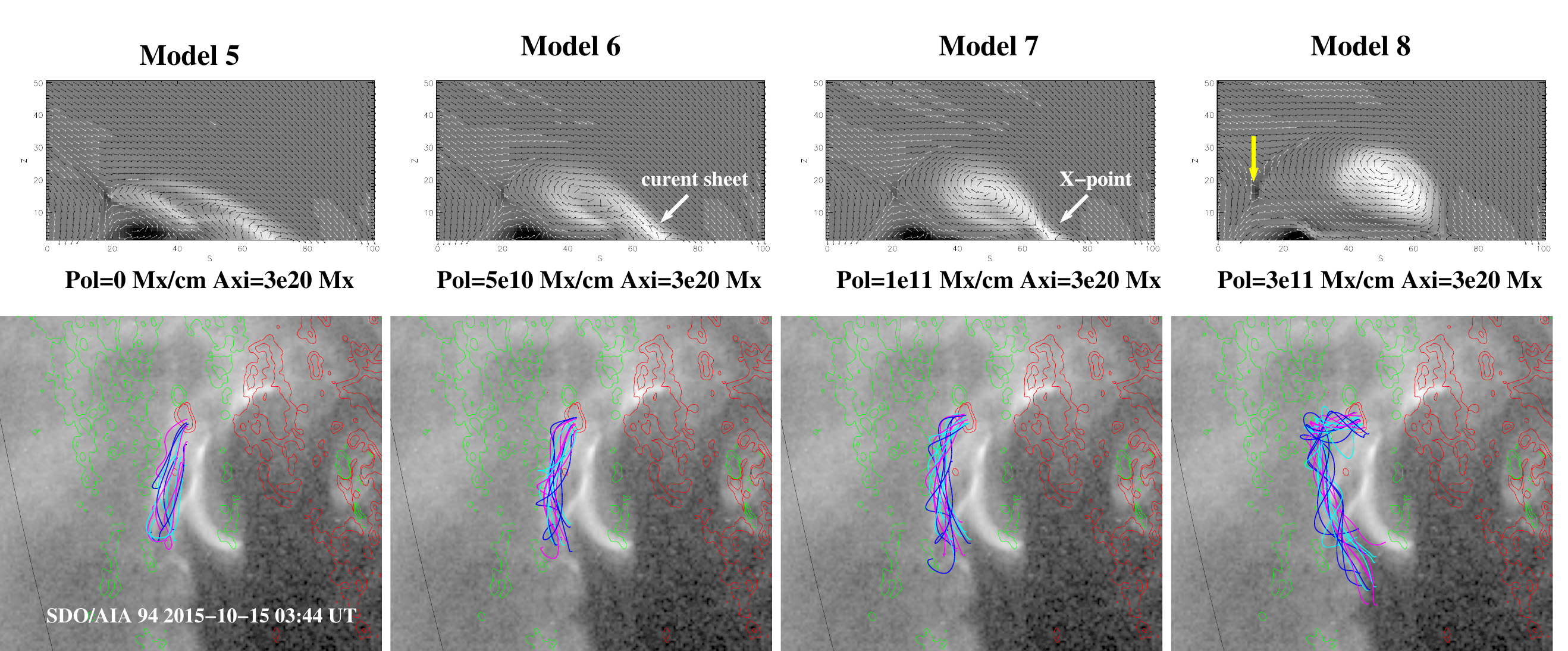}
\caption{Similar to Figure~\ref{fig9}, but the models have fixed axial flux and increasing poloidal flux from left to right. The
maximum value of the current density in the top row is 20000 G $R^{-1}_{\sun}$.
\label{fig10}}
\end{figure}

\clearpage

\begin{figure}[ht!]
\plotone{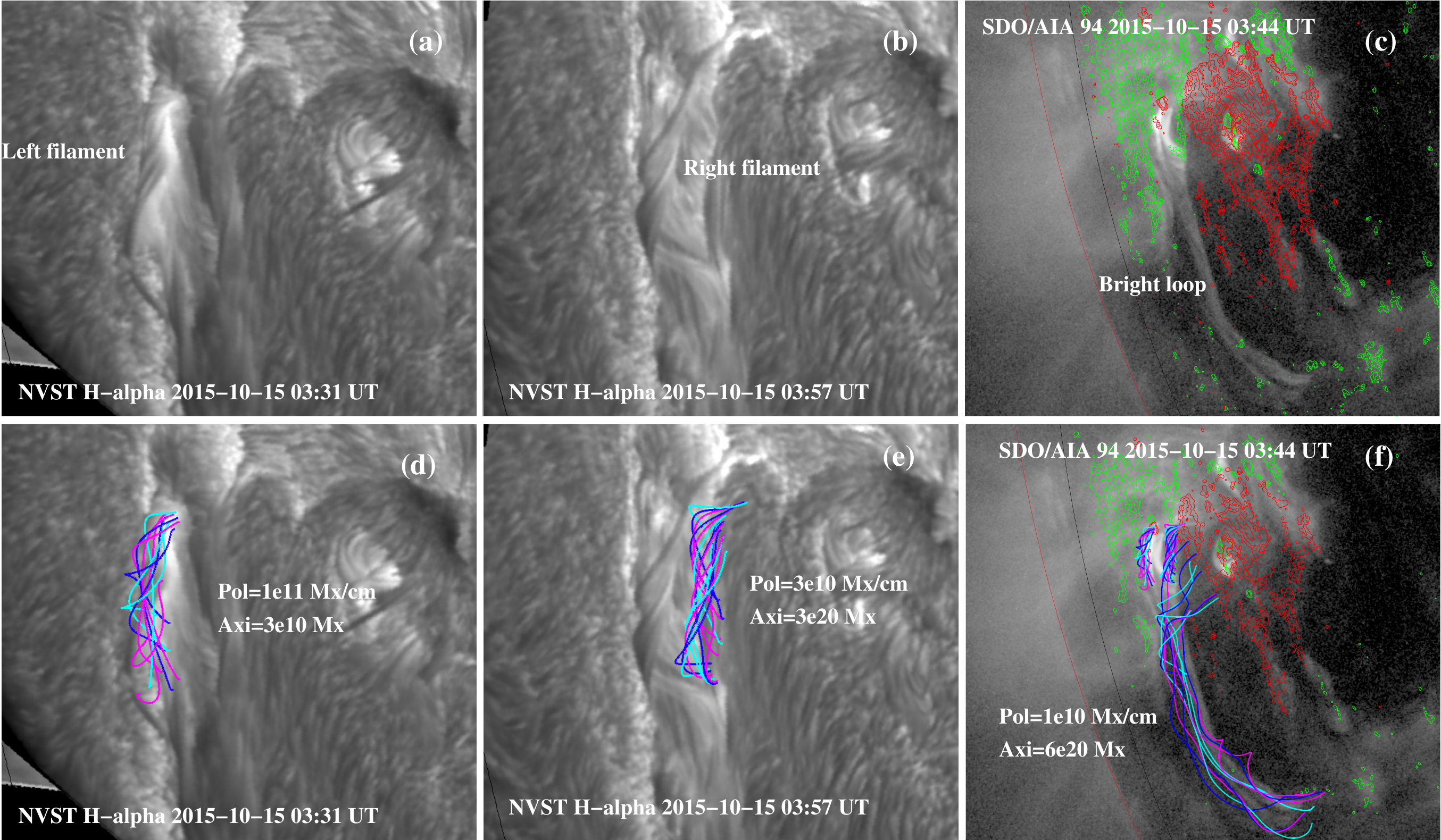}
\caption{Comparison between observed filaments and model flux ropes. (a)--(c) NVST H$\alpha$ images of the left and right filaments with clear twist structure, and AIA 94 {\AA} image with bright long loop. (d)--(f) Selected model field lines representing the left and right filaments, as well as the bright long loop are overlaid on images corresponding to those in the top row.
\label{fig11}}
\end{figure}

\begin{figure}[ht!]
\plotone{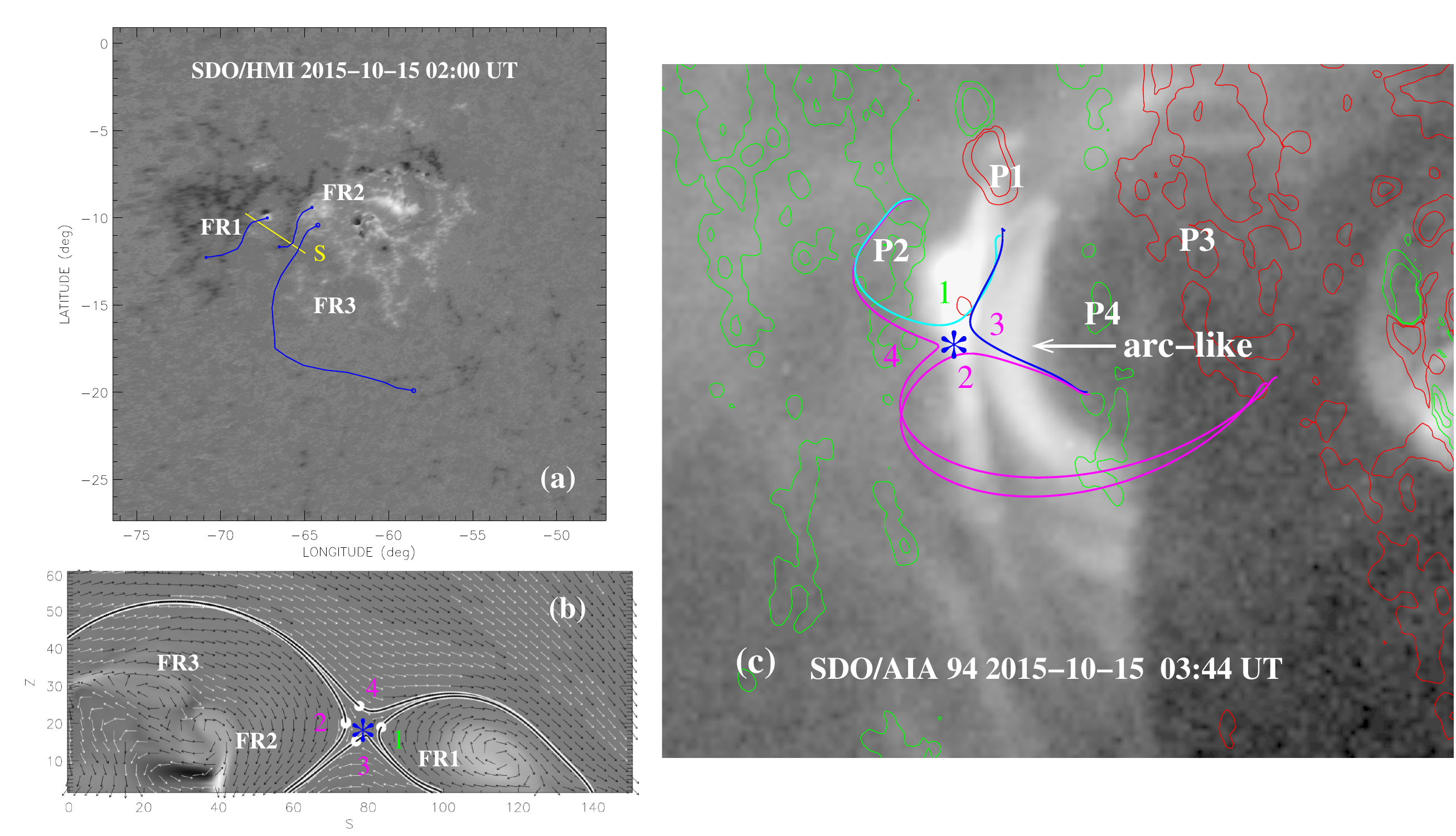}
\caption{Modeling and observation of the bright arc-like structure. (a) The longitude--latitude map of the radial component of the photospheric magnetic field by HMI. (b) Distribution of electric currents in a vertical cross section along the yellow line S marked in (a).  (c) AIA 94~\AA~ image with bright arc-like structure between two filaments. The four white lines in (b) and color lines in (c) refer to selected field lines surrounding the X-point (blue star sign). 
\label{fig12}}
\end{figure}

\clearpage

\end{document}